\def\q{\\}
\def\eotv{E\"otv\"os~}

\documentstyle[aps,prl,multicol,psfig]{revtex}

\begin{document}
\title{Theory of periodic swarming of bacteria: application to Proteus mirabilis}

\author{A. Czir\'ok$^{1,3}$\footnote{czirok@biol-phys.elte.hu}, 
	M. Matsushita$^{2,3}$ and T. Vicsek$^{1,3}$ }

\address{$^1$Department of Biological Physics, \eotv University, 
1117 Budapest, P\'azm\'any stny 1A, Hungary}

\address{$^2$Department of Physics, Chuo University, Kasuga, Bunkyo-ku, Tokyo 112-8551, Japan}

\address{$^3$Institute for Advanced Study, Collegium Budapest, 1014 Budapest, Szenth\'aroms\'ag u 2, Hungary}

\maketitle

\begin{abstract}
The periodic swarming of bacteria is one of the simplest examples for 
pattern formation produced by the self-organized collective behavior of 
a large number of organisms.  In the spectacular colonies of {\em  
Proteus mirabilis} (the most common species exhibiting this type of 
growth) a series of concentric rings are developed as the bacteria 
multiply and swarm following a scenario periodically repeating itself.  
We have developed a theoretical description for this process in order 
to get a deeper insight into some of the typical processes governing 
the phenomena in systems of many interacting living units.

Our approach is based on simple assumptions directly related to the 
latest experimental observations on colony formation under various 
conditions.  The corresponding one-dimensional model 
consists of two coupled differential equations investigated here both by
numerical integrations and by analysing the various expressions obtained
from these equations using a few natural assumptions about the parameters 
of the model.  We have determined the phase diagram corresponding to 
systems exhibiting periodic swarming and discuss in detail how the 
various stages of the colony development can be interpreted in our 
framework.  We point out that all of our theoretical results are in 
excellent agreement with the complete set of available observations.  
Thus, the present study represents one of the few examples, where 
self-organized biological pattern formation is understood within a 
relatively simpe theoretical approach leading to results and predictions 
fully compatible with experiments.
\end{abstract}

\begin{multicols}{2}

\section{Introduction}

To gain an insight into the development and dynamics of various multicellular
assemblies, we must understand how cellular interactions build up the structure
and result in certain functions at the macroscopic, multicellular level.
Microorganism colonies are one of the simplest systems consisting of many
interacting cells and exhibiting a non-trivial macroscopic behavior.
Therefore, a number of recent studies have focussed on experimental and
theoretical aspects of colony formation and the related collective behavior of
microorganisms \cite{SD97,ADD97}.

\begin{figure}

\psfig{figure=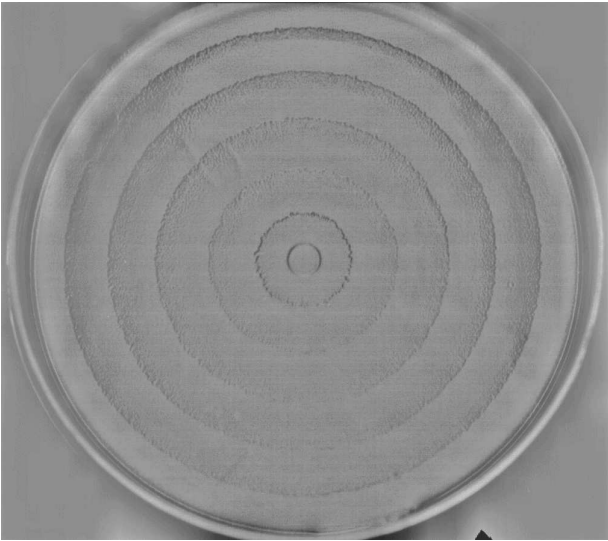,width=8cm}

\caption{
Typical {\it Proteus mirabilis} colony. It was grown on the surface 
   of a 2.0\% agar substrate for two days.  The inner diameter of 
      petri-dish is 8.8 cm. Gray shades are 
proportional to the cell density: the cyclic modulation is apparent.
}
\label{fig1}
\end{figure}

The {\em swarming} cycles exhibited by many bacterial species, notably {\it
Proteus (P.) mirabilis}, have been known for over a century 
\cite{Hauser1885}. 
When {\it Proteus} cells are inoculated on the surface of a
suitable hard agar medium, they grow as short {\em ``vegetative''} rods.
After a certain time, however, cells start to differentiate at the colony
margin into long {\em ``swarmer''} cells possessing up to 50 times more
flagella per unit cell surface area.  These swarmer cells migrate rapidly away
from the colony until they stop and revert by a series of cell fissions into
the vegetative cell form, in a process termed {\em consolidation}. The
resulting vegetative cells grow normally for a time then swarmer cell
differentiation is initiated in the outermost zone (terrace), and the process
continues in periodic cycles resulting in a colony with concentric zonation
depicted in Fig~\ref{fig1}. Similar cyclic behavior has been observed in an
increasing number of Gram-negative and Gram-positive genera including {\it
Proteus, Vibrio, Serratia, Bacillus} and {\it Clostridium} (for reviews see
\cite{WS78,AH91,FH99}). 

The reproducibility and regularity of swarming cycles together with the finding
that its occurrence is not limited to a single species suggest that periodic
swarming phenomena can be understood and quantitatively explained on the basis
of mathematical models.  In this manuscript we first give an overview of
the relevant experimental findings related to the swarming of {\it P.
mirabilis}, then construct a simple model with two limit densities. We then
investigate the behavior of the model as a function of the control parameters
and compare it to experimental results. 

\section{Overview of experimental findings}

\paragraph{Differentiation.}

As reviewed in \cite{AH91,FH99}, the differentiation of vegetative cells is
accompanied by specific biochemical changes.  Swarmer cells enhance the
synthesis of flagellar proteins, extracellular polysaccharides, proteases and
virulence factors, while exhibit reduced overall protein and nucleic acid
synthesis and oxygen uptake. These findings may be explained by arguing that
the production and operation of flagella is expensive and may require the
repression of non-essential biosynthetic pathways.  The largely (10-30 fold)
elongated swarmer cells develop by a specific inhibition of cell fission which
seems {\em not} affecting the doubling time of the cell mass or DNA.

The differentiation process  is initiated by a number of external stimuli
including specific signalling molecules and physico-chemical parameters of the
environment. As an example for the latter, the {\em viscosity} of the
surrounding medium is presumably sensed by the hampered rotation of the
flagella \cite{ALGH93,CS90}. Neither the signal molecules that initiate the
differentiation nor the involved intracellular signalling pathways are
identified yet, but a corresponding transmembrane receptor has been found
recently \cite{BSM98}. The structure of this receptor, together with other
findings reviewed in \cite{FH99}, suggest a `{\em quorum sensing}' regulatory
pathway \cite{FWG96} characteristic for many, cell density dependent collective
bacterial behaviors like sporulation, luminescence, production of antibiotics or
virulence factors\cite{Sha98}.

\paragraph{Migration of swarmer cells.}

It is well established \cite{AH91} that swarmer cell migration does not require
exogenous nutrient sources, since swarmer cells replated onto media devoid of
nutrients continue to migrate normally.  The ability of migration depends on the
local swarmer cell density as isolated single swarmer cells cannot move,
while a group of them can.  It was also demonstrated \cite{WAHSL76} that the
mechanism by which bacteria swarm outwards involved neither repulsive nor
attractive chemotaxis.  The typical swimming velocity (i.e., in liquid
environment) of swarmer cells is $\approx 100 $mm/h \cite{MTNIWM00} which is
also their maximal swarming speed and rate of colony expansion on soft agar
plates \cite{AH91}.  In the usual experimental conditions for investigating
swarming colony formation the front advances with a speed of $0.5-10$ mm/h
\cite{RMWSES96,BSM98,IWMM99,MTNIWM00}.  Unfortunately, in these cases there is
no information available on the velocity of individual swarmer cells, but it 
must be between the colony expansion speed and the swimming velocity.

\paragraph{Consolidation.}

The molecular mechanisms of consolidation, i.e., the downregulation of the gene
activity responsible for swarming behavior \cite{DFH98} is even less known than
that of differentiation. If the swarming motility utilizes intracellular energy
reserves as has been suggested \cite{AH91}, then swarmer cells must have a {\em
finite lifetime}. In addition to the septation of swarmer cells taking place at
the outermost terrace, inside the colony the differentiation process, i.e., the
supply of fresh swarmer cells must also be shut off. The
cessation of swarmer cell production does not seem to be due to severe nutrient
depletion since vegetative cells keep growing (although with decreasing growth
rate) well inside the colony for many hours after the last swarmer cells were
produced in that region \cite{RMWSES96}.

\paragraph{Colony formation.}

The cycle time (total length of the migration and consolidation periods) has
been found \cite{RMWSES96} to be rather stable ($\approx 4h$) for a wide
range of nutrient and agar content of the medium.  The size of the terraces and
the duration of the migration phases were strongly influenced (up to an order of
magnitude, and up to a factor of 3, respectively) by the agar hardness. The
nutrient concentration did not have an observable effect on these quantities
from $0.01\%$ up to $1\%$ glucose concentrations. There was, however a
remarkable positive correlation between the cycle time and the doubling time
(ranging from $0.7$h up to $1.8$h) of the cells.

Two interfacing colonies inoculated with a time difference of a few hours and
therefore being in different phases of the migration-consolidation cycle, were
found to maintain their characteristic phases \cite{RMWSES96,IWMM99}. Thus, the
control of the swarming cycle must be sufficiently local.

\paragraph{Cycle rescheduling.}

A few experiments investigated the cell density dependence of the duration of
quiescent growth ({\em lag phase}) prior to the first migration phase.  These
studies clearly revealed that vegetative cells have to reach a threshold density
to initiate swarming \cite{RMWSES96,IWMM99}, in accord with the suggested
quorum sensing molecular pathway of the initiation of swarmer cell 
differentiation.

Agar cutting experiments demonstrated that a cut inside the inner terraces does
not influence the swarming activity \cite{IWMM99}.  However, when the cut has
been made just behind the swarming front, the duration of the swarming phase
was shortened and consolidation was lengthened by up to $40\%$ \cite{IWMM99}.

Even more interestingly, {\em mechanical mixing} of the cell populations before
the expected beginning of consolidation expands the duration of the swarming
phase considerably, by up to $50\%$\cite{MTNIWM00}. This finding, together with
{\em replica-printing} experiments\cite{MTNIWM00} demonstrates that at the
beginning of the consolidation phase still a large pool of swarmer cells exists
and seems to be ``trapped'' at the rear of the outermost terrace.

\section{The model}

\def\rs{\rho}
\def\rso{\rho^o}
\def\rw{\rho^*}
\def\Gs{\Gamma}
\def\Gw{\Gamma^*}
\def\rd{\tilde{\rho}}

\def\be{\begin{equation}}
\def\ee{\end{equation}}
\def\bea{\begin{eqnarray}}
\def\eea{\end{eqnarray}}

Taking into account the above described experimental findings, here we
construct a model which is capable of explaining most of the observed features
of colony expansion through swarming cycles.

\noindent (i) 
The model is based on the vegetative and swarmer cell population densities {\em
only}, denoted by $\rso$ and $\rw$, respectively. These values are defined on
the basis of cell mass instead of cell number, therefore one unit of swarmer
cells is transformed into one unit of vegetative cells during consolidation.

\noindent (ii) 
Vegetative cells grow and divide with a constant rate $r_0\approx
1$h$^{-1}$ \cite{RMWSES96,IWMM99}. This will later allow us to establish 
a direct correspondence between cell density increase and elapsed time. 

\noindent (iii) 
Usually, swarmer cell differentiation is initiated when the local density of
the vegetative cells exceeds a threshold value ($\rso_{min}\approx 10^{-2}$
cells/$\mu$m$^2$ \cite{RMWSES96,IWMM99}). (Prior the first swarming phase
experiments indicate the presence of an extra time period $t_\ell$ which is
probably associated with the biochemical changes required to develop the
ability of the swarming transition. This effect is present only at the
seeding of the colony, thus $t_\ell=0$ otherwise.) When
$\rso=\rso_{min}$ at time $t_0$, some of the vegetative cells enter the
differentiation process, modeled by introducing a rate $r$.  Since the biomass
production rate is assumed to be unchanged during the differentiation process,
the rate of producing new vegetative cells is $r_0-r$ and the differentiating
cells elongate with the normal growth rate $r_0$.

\noindent (iv)
The full development of swarmer cells, i.e., a typical 20-fold increase in
length needs a time ($t_d\approx\ln20/r_0\approx3$h) comparable with, or even
longer than the length of a consolidation period, hence can not be neglected.
Therefore, the first ``real'' swarmer cells, which are able to move appear only
at $t_0+t_d$.  

\noindent (v) The production of swarmer cells is limited in time and
the length $\tau$ of the time interval during which vegetative cells can enter
the differentiation process is another phenomenological parameter of our model. 

As we here focus on the periodicity of colony expansion, we do not consider
what happens in the densely populated regions after $t_0 + \tau$.
Specifically, in our model any activity of the vegetative cells ceases in these
parts of the system.

\noindent (vi) Swarmer cells can migrate only if their density exceeds a
threshold density $\rw_{min}$. Above that threshold, swarmer cells are assumed
to move randomly with a diffusion constant $D_0$.  The finite lifetime of
swarmer cells is incorporated into the model through a constant rate ($r^*$)
decay. 

Unfortunately there are no good estimates on these parameters in the 
literature. According to recent experimental observations \cite{Matsu},
$\rw_{min}$ is less than $10\%$ of the value of $\rso$ prior the beginning 
of the swarming phase, i.e., $0.1\rso_{min}e^{r_0t_d}\approx10^{-2}$
cells/$\mu$m$^2$.  $D_0$ may be estimated as $v_0^2t_p/2$ with $v_0$ being the
typical speed of swarming cells and $t_p$ being the {\em persistence time} of
their motion. As $v_0$ is approximately $30$mm/h (see Sec II.b)  and $t_p$ is
in the order of minutes, $D_0$ is estimated to be in the order of 10 mm$^2$/h.

The above considerations lead to the following set of equations
\bea
\dot\rso(t) &=& r_0\rso(t) + \Gamma(t) - \Gamma^*(t) \cr
\dot\rw(t) &=& - \Gamma(t) + \Gamma^*(t-t_d)e^{r_0t_d} +\nabla D(\rw) \nabla \rw
\label{M1}
\eea
where the consolidation ($\Gamma$) and differentiation ($\Gamma^*$) terms are
given by
\be
\Gamma=r^*\rw 
\hbox{~~and~~}
\Gamma^*=r(\rso,t)\rso.
\ee

\begin{figure}

\psfig{figure=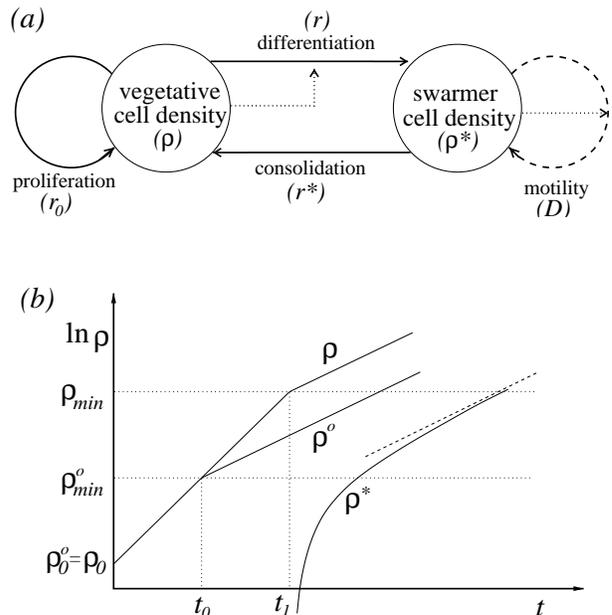,width=8cm}

\caption{
Schematic representation of the model for swarming colony formation.
(a) The two basic quantities are the vegetative and swarmer cell densities, 
which 
can be transformed into each other, and changed by proliferation and motility.
The dotted lines represent regulatory (threshold) effects: The rate of 
differentiation is assumed to be dependent on the vegetative cell density, and
the motility of swarmer cells is also determined by their local density.
(b) Notations and typical time courses of vegetative and swarmer cell densities 
prior the beginning of the first migration phase. If $t<t_0$ the density
$\rho$ (or $\rho^o$) grows with a constant rate $r_0$. After reaching
the density threshold $\rho^o_{min}$, vegetative cells keep growing only with a
rate $r_0-r$, but the total density ($\rho$) of the vegetative cells and the 
differentiating  swarmer cells still grows with a rate $r_0$. After the time
required for the full elongation of a swarmer cell (when $\rho$ reaches 
$\rho_{min}$), $\rho^*$ becomes positive, and for $r^*=0$ asymptotically would
grow with a rate $r$.
}
\label{fig2}
\end{figure}

Eqs.~(\ref{M1}) can be significantly further simplified by making use of the 
possibility to measure elapsed time with the increase in
$\rso$. In particular, neglecting correction terms related to consolidation 
in areas where $\rso>\rso_{min}$, i.e.,
where primarily differentiation takes place, we can cast 
$\Gamma^*(t-t_d)e^{r_0t_d}$ in the form 
\bea
r\rso(t-t_d)e^{r_0t_d}=
r\rso_{min}e^{(r_0-r)(t-t_d-t_0)}e^{r_0t_d}=\cr
r\rso_{min}e^{r_0(t-t_0)}e^{-r(t-t_0-t_d)}=
r\rso(t)e^{rt_d}
\label{tmp}
\eea
for $t>t_0+t_d\equiv t_1$. Let us introduce a transformed population density 
$\rs$ as
\be
\rs(t)= \left\{
\begin{array}{ccl}
  \rso(t) &	\mbox{for~~} \rso<\rso_{min} & \mbox{i.e., for~~} t<t_0 \cr
\rso(t)e^{r(t-t_0)}&\mbox{for~~} \rso>\rso_{0} &	\mbox{and~~} t<t_1 \cr
\rso(t)e^{rt_d} &\mbox{for~~} \rso>\rso_{0} &	\mbox{and~~} t>t_1
\end{array} \right.  
\ee
which is in fact the total density of the vegetative and the differentiating,
but not yet fully differentiated swarmer cells (see Fig.~\ref{fig2}). With this
notation, using (\ref{tmp}) and similar considerations for $t_0<t<t_0+t_d$,
Eqs.~(\ref{M1}) can be written into a simple, not retarded form 
\bea
\dot\rs &=& r_0\rs + r^*\rw - r(\rs)\rs \cr
\dot\rw &=& - r^*\rw + r(\rs)\rs +\nabla D(\rw) \nabla \rw,
\label{M2}
\eea
where
\be
r(\rs) = \left\{
\begin{array}{cl}
  r  &  \mbox{for~~} \rs_{min}<\rs<\rs_{max} \cr
  0  &	\mbox{otherwise}
\end{array} \right.  
\ee
with $\rs_{min}=e^{rt_d}\rs_{min} \approx 10^{-1}$ 
cells/$\mu$m$^2$ and $\rs_{max}(\tau)=e^{(r_0-r)\tau}\rs_{min}$. 
If the $\Gamma \ll (r_0-r)\rs$ condition does not hold in the $[t_0, t_1]$
time interval, then $\rs_{min}$ must be also treated as a 
dynamical variable. This case will not be considered here.
In the following we use $\rs$ for the characterization of
vegetative cell density.

\section{Results}

\subsection{Numerical method in 1D} 

The model defined through  Eqs.~(\ref{M2}) has the following seven parameters:
the rates $r_0$, $r$, $r^*$, threshold densities $\rho_{min}$,
$\rho_{max}$ (or $\tau$), $\rho^*_{min}$ and the diffusivity $D_0$. 
However, this number can be reduced to four by casting the
equations in a dimensionless form using $1/r_0\approx1$h as time
unit, $\rho_{min}\approx0.1$ cells/$\mu$m$^2$ as density unit and
$x_0=\sqrt{D_0/r_0}\approx3$mm as the unit length. The resulting control
parameters are $r/r_0$, $r^*/r_0$, $\rs_{max}/\rs_{min}=\exp[(r_0-r)\tau]$
and $\rw_{min}/\rs_{min}$.
To obtain continuous density profiles, the step-function dependence of
$D$ on $\rw$ was replaced by

\be
D(\rw)={D_0\over2}\left[ 1 + \tanh2\alpha{\rw-\rw_{min}\over\rw_{min}}\right]
\label{nonlinD}
\ee
with $\alpha=10$ providing a rather steep, but continuous crossover.

\begin{figure}

\psfig{figure=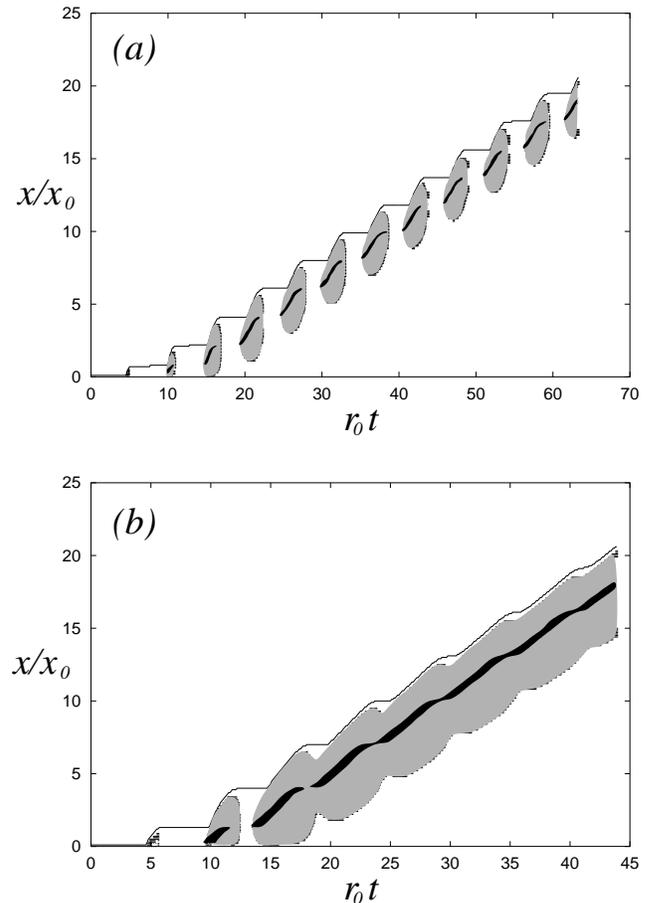,width=8.5cm}

\caption{
Time development of the model obtained by numerical integration of the
equations, starting from a localized  ``inoculum'' at $t=0$, $x=0$.  
The continuous line
represent the colony boundary (maximal value of $x$ for which
$\rho(x)+\rho^*(x)>0$). The filled gray and black areas  are
regions where swarmer cells are motile, and where swarmer cells are
produced, respectively. For $\rho_{max}=1.3$, $\rho^*_{min}=0.01$,
$r=0.3$ and $r^*=1.0$ the expansion of the system is clearly periodic (a). If
we increase the production of swarmer cells by increasing $\rho_{max}$ to
$2.0$ then the periodicity is gradually lost and a continuous expansion takes
place (b).
} 
\label{fig3}
\end{figure}

Representative examples for the time development of the model are shown in 
Fig.~\ref{fig3}.
The production of swarmer cells is localized, and determined
by the density profile of vegetative cells at the end of migration periods.
In this particular model $\rs(x)$ is decreasing towards the colony edge,
therefore in the migration phases the source of swarmer cells is moving
outwards. The front of swarmer cells is
expanding from the inside of the last terrace. Because of the decay term 
$\Gamma$, cells become non-motile first at the colony edge.

\subsection{Phase diagram}

Each of the dimensionless control parameters can have an important effect on
the dynamics of the system. As an example, if the duration $\tau$ of swarmer
cell production is increased, then the consecutive swarming cycles are not
separated and a continuous expansion takes place with damped oscillations
(Fig.~\ref{fig3}b).  To map the behavior of the system as a function of the
control parameters, the following procedure was applied.  Migration periods
were identifed by requiring $\max_x\rw(x)>\rw_{min}$.  For a given set of
parameters we determined the lengths $\{t_i\}$ of the consecutive migration
periods, and the system was classified as {\em periodic} if the three largest
values of the set $\{t_i\}$ were the same within $20\%$. Otherwise, the
expansion was classified as {\em continuous} as long as $\max_i\{t_i\}$ was
large enough: comparable with the total duration of the simulated expansion.

\begin{figure}

\psfig{figure=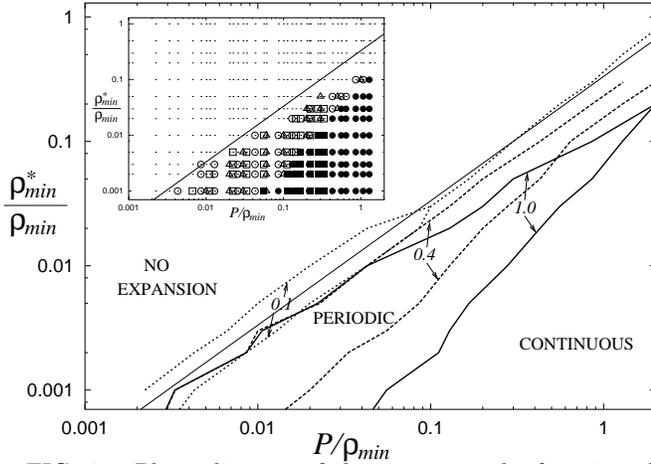,width=9cm,angle=-90}

\caption{
Phase diagram of the system as the function of cumulative swarmer cell 
production density $P$ and migration threshold $\rho^*_{min}$ for various
values of swarmer cell decay rate $r^*$. If the production is high or the
motility threshold is low enough then continuous expansion can be observed.
On the other hand, if the production is too low or the motility threshold is 
too high then no expansion takes place. In an intermediate regime periodic
growth can be observed. The boundaries of this parameter regime are
plotted for $r^*/r_0=1.0$ (thick continuous line), $0.4$ (thick dashed line)
and $0.1$ (thick dotted line).  The thin continuous line represent an
approximate upper bound (\ref{approx_phase}) for cyclic colony expansion.
The insert demonstrates that the fourth parameter of the model, $r$, is 
irrelevant: for $r^*/r_0=1$ and $r/r_0=0.01$ ($\Box$), $0.3$ ($\circ$), $0.5$
($\triangle$) and various values of $\rho_{max}$ and $\rho^*_{min}$
the type of colony expansion was classified. Open symbols correspond to
cyclic growth, filled symbols to continuous growth while dots denote
no expansion. Note that the corresponding regions completely overlap
irrespectively of the value of $r$.
}
\label{fig4}
\end{figure}

The behavior of the model is summarized in Fig.~\ref{fig4},
where the boundaries of the various regimes are plotted for three
different values of $r^*/r_0$. We found, that $r$ and $\rs_{max}$
can be combined into one relevant parameter, the swarmer cell
production density, as
\be
P=\int_{-\infty}^\infty \Gamma^*(x,t)dt=
{r\over r_0-r}(\rs_{max}-\rs_{min}),
\ee
which quantity does not depend on the choice of position $x$.  As the
insert demonstrates, for a given $P$, the actual values of $r$ or
$\rs_{max}$ are irrelevant to this kind of classification in the parameter
regime investigated.  The general structure of the phase diagram was found to
be similar for various values of $r^*$. For large enough $P$ or low enough
$\rw_{min}$ a continuous expansion takes place, while for too small $P$ or
large $\rw_{min}$ the  expansion of the system is finite. For intermediate
values of these parameters an oscillating growth develops exhibiting well
distinguishable consolidation and migration phases.  As the lifetime of the
swarmer cells is increased, the parameter regime, in which periodic behavior is
exhibited, is shrinked and moved towards lower $P$ values.

One can easily estimate the position of the boundary of the non growing
phase based on that (i) the width $w$ of the terraces is small (this assumption
is justified later, in Fig.~\ref{fig7}.), thus (ii) the time required for the diffusive
expansion of the swarmer cells is much shorter than their lifetime,
which, in turn, is (iii) shorter than
the duration of a swarming cycle: $r^*/r_0\sim1$. 
The amount of swarmer cells produced in one period is $Pw$. Neglecting
the decay during expansion, the width $w'$ of the next, new terrace can be
determined from the conservation of cell number as
\be
2\rw_{min}w'=w(P+\rw_{r}-\rw_{min}),
\ee
where $\rw_{r}$ denotes the swarmer cell density remaining from the previous
swarming cycle and the symmetric expansion of the released swarmers was also
taken into account. To achieve a sustainable growth $w'\geq w$ is required,
resulting in  a condition 
$3\rw_{min}\leq P+\rw_{r}$.
If $\rw_{r}\ll P$, as one can expect for $r^*/r_0\sim1$, we get for the
boundary of the non growing phase
\be
P=3\rw_{min},
\label{approx_phase}
\ee 
which,  as Fig.~\ref{fig4} demonstrates, is indeed in
good agreement with the numerical data.

\subsection{Terrace formation.}

\begin{figure}

\centerline{\psfig{figure=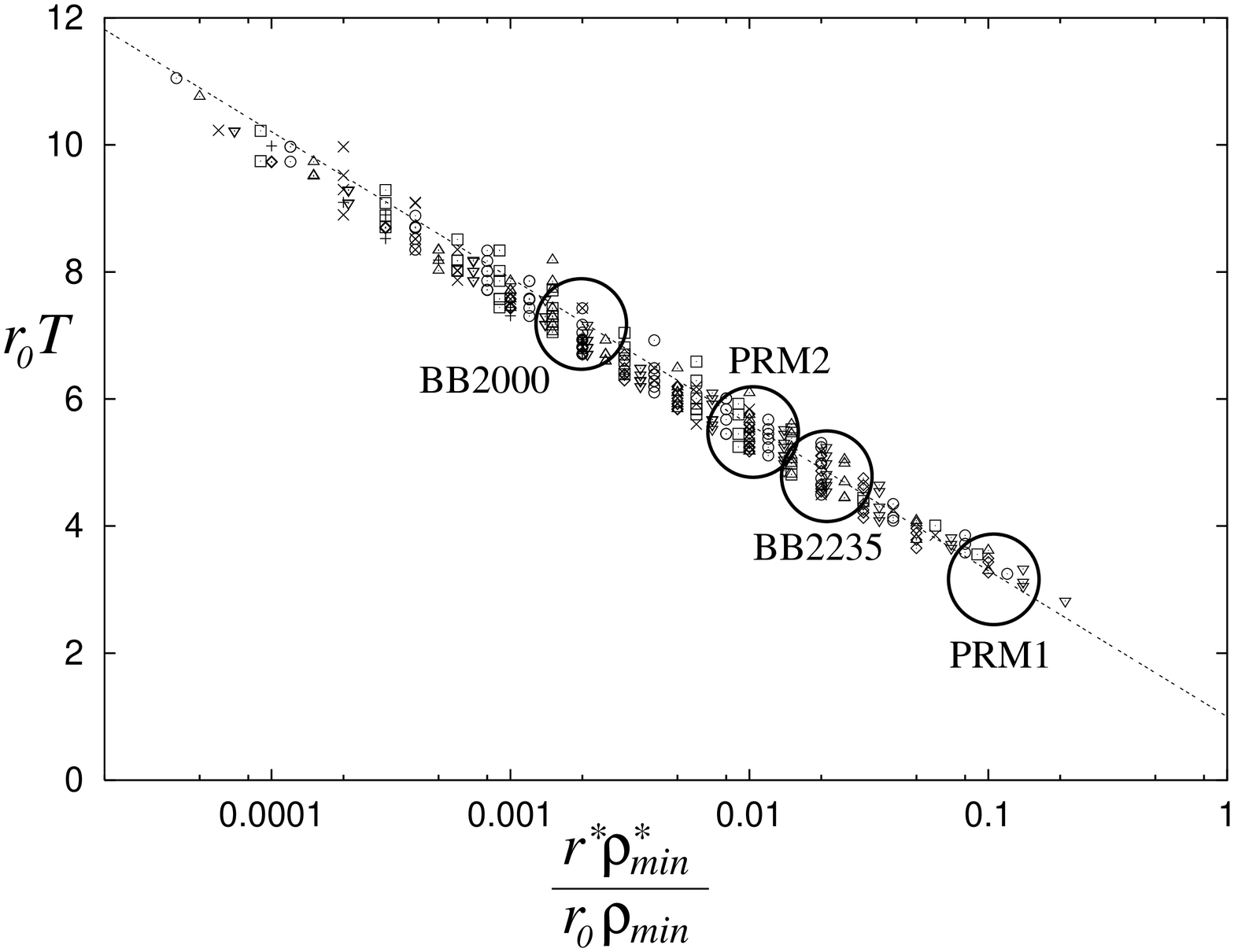,width=8.0cm,angle=-90}}

\caption{
The cycle time $T$ as a function of the approximate consolidation rate
$r^*\rho^*_{min}$. The data collapse indicates that the impact on $T$ of the
other parameters ($\rho_{max}$ and $r$) is negligible. The various symbols
correspond to different values of $r^*/r_0$ as $0.1$ ($+$), $0.2$ ($\times$),
$0.3$ ($\Box$), $0.4$ ($\circ$), $0.5$ ($\triangle$), $0.7$ ($\bigtriangledown$), 
$1.0$ ($\diamond$). The dashed line is a plot of Eq.(\ref{resT}). 
Circles mark out the assumed parameter values characteristic of four 
different {\it P. mirabilis} strains.
}
\label{fig5}
\end{figure}

The average length $T$ of a full swarming cycle was calculated by determining
the position of the peak in the power spectrum of $S(t)=\int_0^\infty
\rw(x,t)dx$, the time dependence of the total number of swarmer cells in
the system.  As Fig.~\ref{fig5} demonstrates, the dimensionless cycle time
values are widely spread between values of $3$ and $12$. However, $T$ is {\em
only} sensitive to changes in $r_0$, $r^*$ and $\rw_{min}/\rs_{min}$, hence it
does not depend on $\rs_{max}$ or $\tau$.

\begin{figure}

\centerline{\psfig{figure=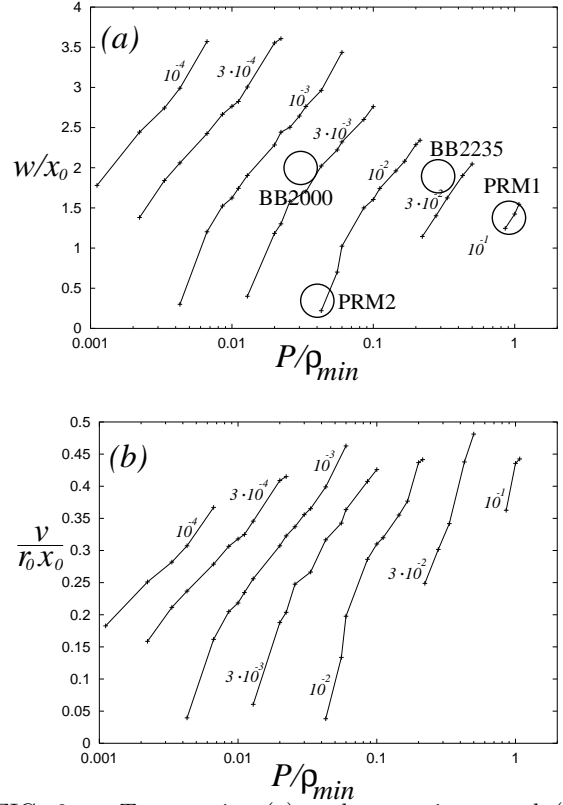,width=7.50cm}}

\caption{
Terrace size (a) and expansion speed (b) of the system for $r^*/r_0=1$ as a function
of the dimensionless swarmer cell production density $P$. The connected points
correspond to various values of the migration threshold 
$\rho_{min}^*/\rho_{min}$.
In general, decreasing $\rho_{min}^*$ or increasing $P$ results in an 
increase in both $w$ and $v$. Only parameters resulting in periodic expansion
were investigated. Circles mark out the assumed parameter values characteristic of four
different {\it P. mirabilis} strains.
}
\label{fig6}
\end{figure}

\begin{figure}

\psfig{figure=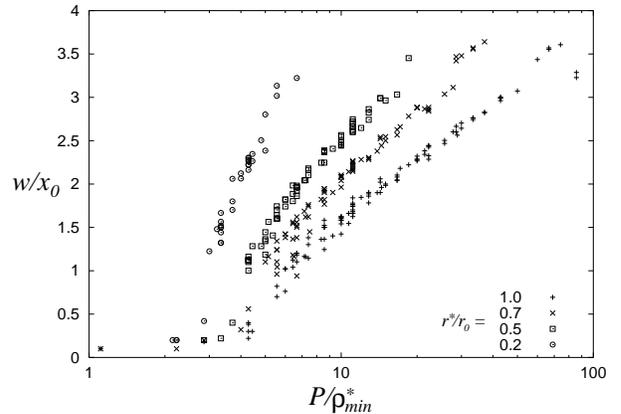,width=8cm,angle=-90}

\caption{
Terrace size $w$ vs $P/\rho_{min}^*$ for various values of $r^*/r_0$.
Note the collapse of the data presented in Fig.~6a. The terrace size
vanishes approaching the parameter regime where no sustainable expansion of the 
system is possible.
}
\label{fig7}
\end{figure}

The average expansion speed $v$ and terrace size $w$ were also calculated 
in the parameter regime resulting oscillatory expansion of the colony.
First we determined the time $t_{1/3}$ when the system reached $1/3$ of its
maximal simulated expansion $R_{max}=R(t_{max})$, with $R(t)$ being the
position of the expanding colony edge and $t_{max}$ is the total duration of
the simulation. The average speed was then calculated for the time interval
between $t_{min}=\max(t_{1/3}, t_{max}-5T)$ and $t_{max}$: for the last
$5T$ long time interval, or for the last $2/3$rd of the total expansion,
depending on which was smaller. After obtaining $v$ as
$[R_{max}-R(t_{min})]/(t_{max}-t_{min})$, the average terrace width was
calculated as $w=vT$.  Fig.~\ref{fig6}. shows the dependence of these parameters
on the swarmer cell production density $P$ and migration density threshold
$\rw_{min}$.  In general, decreasing $\rho_{min}^*$ or increasing $P$ results
in an increase in both $w$ and $v$.  As Fig.~\ref{fig7}. demonstrates, for a given $r^*$,
the relevant parameter controlling $w$ is $P/\rho_{min}^*$.

The results on the cycle time $T$ (Fig.~\ref{fig5}.) can be interpreted as follows.  As
$\rw_{min}$ is a good estimate on the density of swarmer cells in the expanding
front, at the end of migration phase the vegetative cell density within the 
new terrace is given by $r^*\rw_{min}T^*$ with $T^*$ being the
duration of the migration phase. Now
the length of the consolidation phase, $T-T^*$, is determined by
the requirement that $\rs$ must reach $\rs_{min}$:
\be
\rs_{min}=r^*\rw_{min}T^*e^{r_0(T-T^*)}=
r^*\rw_{min}{T^*\over e^{r_0T^*}}e^{r_0T}.
\label{est-1}
\ee
As $T^*\approx 1/r_0$, the estimate (\ref{est-1}) is simplified to
\be
r_0 T = \ln {r_0\rs_{min}e\over r^*\rw_{min}},
\label{resT}
\ee
which gives a rather accurate fit to the numerically determined data (Fig.~\ref{fig5}).

\subsection{Lag phase}

\begin{figure}

\psfig{figure=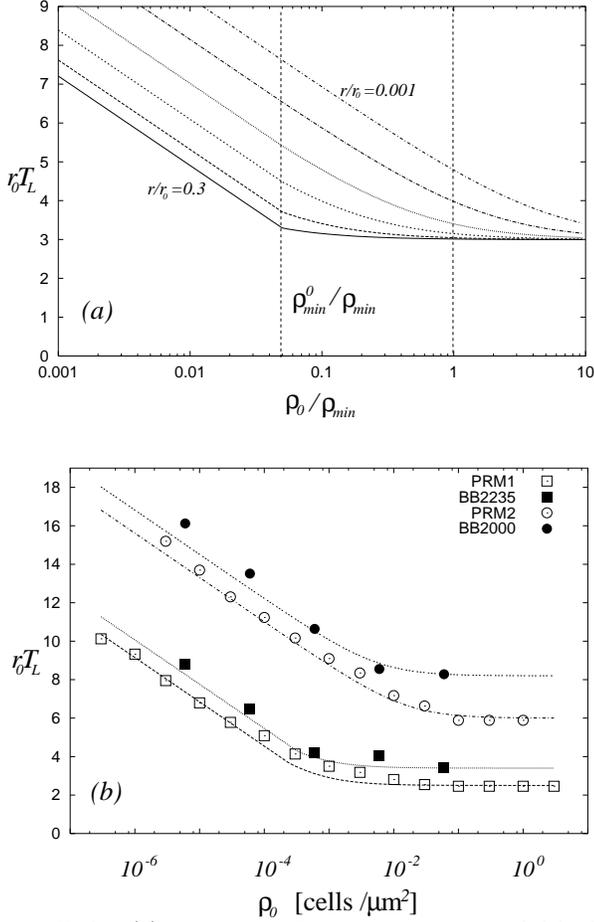,width=8cm}

\caption{
(a) Length $T_L$ of the lag phase vs the initial inoculum density $\rho_0$
for $t_\ell=0$, $r_0t_d=3$ and various values of $r/r_0$ (0.3, 0.1,
0.03, 0.01, 0.003 and 0.001). The transition between the regimes
separated by $\rho^o_{min}$ becomes more smooth for decreasing $r/r_0$.
(b) The experimentally calculated $r_0T_L$ values vs the estimated inoculum 
density (based on 5 mm and 7 mm inoculum droplet sizes for the RPM and BB
strains, respectively) and the corresponding fits using Eqs. (\ref{resL0}) 
and (\ref{resL2}).
}
\label{fig8}
\end{figure}

Since the duration ($T_L$) of the lag phase (the time period before the first
migration phase) has been in the
focus of many recent experiments, now we turn our attention towards
this quantity. At least four processes determine $T_L$. First, there is a time
$t_\ell$ associated with the biochemical changes required to switch into
swarming mode. As discussed in Sec.  II.a, these processes take place only
prior the first swarming phase, and are presumably related to sensing the
altered environmental conditions. Second, the cell population must reach the
threshold density $\rso_{min}$ (at time $t_0$). Third, $t_d$ time is required to
produce fully differentiated swarmer cells (at time $t_1$), and finally, the
density of the swarmer cells must reach the migration threshold $\rw_{min}$.
Let us investigate how these parameters depend on the initial inoculum density
$\rs_0$.

As $\rho$ grows with a rate $r_0$ until the appearance of swarmer cells,
\be
r_0t_1=\max\left[ t_\ell + t_d, \ln(\rs_{min}/\rs_0) \right].
\label{resL0}
\ee
The time development of 
$\rw$ can be estimated by the integration of Eq.(\ref{M2}) with $r^*=0$
(i.e., assuming $\Gamma\ll\Gamma^*$) yielding
\be
\rw(t+t_1)={r\over r_0-r}\rs_0e^{r_0t_1}[e^{(r_0-r)t}-1].
\label{resL1}
\ee
Therefore, $t^*=T_L-t_1$ is given by
\be
r_0 t^* = 
{1\over1-r/r_0}\ln\left({r_0-r\over r}{\rw_{min}\over\rs_0e^{r_0t_1}} + 1\right),
\label{resL2}
\ee
an expression usually giving a minor correction to $t_1$.

Fig.~\ref{fig8}a. shows the above calculated $r_0T_L$ vs $\rs_0$
for $t_\ell=0$. The increase in length of the swarmer cells was assumed to be
$20$-fold, thus $r_0t_d=\ln20\approx3$, which value can be seen for
$\rs_0\gg\rs_{min}$.  In the opposite limit, when  $\rs_0\ll\rs_{min}$, we have
$r_0T_L\approx -\ln\rs_0 + const$.  These relations allow the determination of
both $r_0$ and $r$ (using the known value of $\rs_{min}/\rw_{min}$) from the 
experimental data on $T_L(\rs_0)$. 

\subsection{Comparison with experiments}

Most of the published experimental data are related to the average period
length, $T$ and terrace size $w$.  From these parameters the average expansion
speed can be calculated as $v=w/T$, i.e., $v$ is not an independent quantity.
As we could see in the previous paragraph, from the density dependence of the
lag phase the parameters $r_0$, $t_\ell+t_d$ and $r$ can be estimated.  Notice
that this estimate on $r_0$ is in principle different from the value obtained
by the usual methods based on densitometry in liquid cultures.  Technically,
$\rw_{min}$ could be also determined \cite{RMWSES96}, but such measurements are
not published yet.

There are four {\it Proteus} strains studied systematically in experiments: 
the PRM1, PRM2, BB2000 and BB2235 strains (see Table I). 
To extract the values of the
model's parameters the following procedure was applied. (i) We estimated $r_0$
based on lag phase length measurements. (ii) From the calculated $r_0T$ values
the $\rw_{min}/\rs_{min}$ ratio was estimated (assuming $r^*/r_0=1$)
based on Eq.(\ref{resT}),
see Fig.~\ref{fig5}. (iii) Using Eqs. (\ref{resL1}) and
(\ref{resL2}), by a nonlinear fitting procedure (Levenberg-Marquardt method,
\cite{numrec}, see Fig.~\ref{fig8}b.) $\rs_{min}$, $r$, and $t_\ell+t_d$ was
determined. (The latter value is not relevant in respect the periodicity of the
behavior.) (iv) Knowing $\rw_{min}$ and $\rs_{min}$, from the experimental
terrace width data $x_0$ and $P$ can be estimated using Fig.~\ref{fig6}.  (v)
Finally, $\tau$ is given by
\be
r_0\tau=\ln\left({Pr_0\over\rs_{min}r}+1\right)
\ee
The parameter values of the model are summarized in Table II, together with 
the predictions on $T$, $v$ and $w$.  An excellent agreement
can be achieved with biologically relevant parameter values.

Two classes of model parameters should be distinguished: (a) the ones which are
related to the growth and differentiation of the cells ($r_0$, $r^*$,
$\rs_{min}$, and $P(r,\tau)$) and (b) those which depend on agar softness
($D_0$ and $\rw_{min}$). For a given strain we expect that a change in the agar
concentration influences only the latter group, while changes in temperature may
affect both, but primarily $r_0$. In fact, as Table I. demonstrates, by
changing $D_0$ and keeping all the other, growth-related parameters constant,
we could {\em quantitatively} reproduce the colony behavior observed on various
agar concentrations. Similar statement holds for the temperature effects as 
well, where the only parameter we changed was the growth rate $r_0$.

\section{Discussion}

Periodic bacterial growth  patterns have been in the focus of research
in the last few years. Since a colony can be viewed as a 
system where diffusing nutrients are converted into
diffusing bacteria, one may not be surprised by the emergence of
spatial structures \cite{Cross93}. However, the periodic patterns
of bacterial colonies are qualitatively different from 
the Liesegang rings (for a recent review see \cite{Antal98}) 
developing in reaction-diffusion systems: the spacing between the densely
populated areas is uniform and independent of the concentration of the 
other diffusing species, i.e., the nutrients. The Turing instability is also
well-known for producing spatial structures \cite{Murray}, but in that case
the pattern emerges simultaneously in the whole system. It is also known,
that bacteria can aggregate in steady concentric ring structures as a 
consequence of chemotactic interactions \cite{Tsimring95,Woodward95},
but as we discussed in Sec. II., it is established that swarming of {\it P. 
Mirabilis} does not involve chemotaxis communication.
Therefore, none of the well known generic pattern forming schemes can explain
the colony structure of swarming bacteria.

As we mentioned in the introduction, oscillatory growth is also exhibited 
by other bacterial species. One of them, {\em Bacillus subtilis}, has been the 
subject of systematic studies on colony formation and a number of models
have been constructed to explain the observed morphology diagram 
(for recent reviews
see \cite{Golding98,Matsu99}). Only one model addressed the problem of
migration and consolidation phases: Mimura et al \cite{Mimura00} 
set up a reaction-diffusion system in which the decay rate of the bacteria was
dependent both on their concentration and the locally available amount of
nutrients. The periodic behavior is then a consequence of the following cycle:
if nutrients are used up locally, then the bacterial density starts to decay
preventing the further expansion of the colony. Nutrients diffuse to the colony
and accumulate due to the reduced consumption of the already decreased
population. The increased nutrient concentration gradually allows the increase
in population density and the expansion of the colony, which starts the cycle
from the beginning. While this can be a sound explanation for {\it B. subtilis},
as we discussed in Sec. II., the nutrient limitation clearly can not explain 
neither the differentiation nor the consolidation of {\it
P. Mirabilis} swarmer cells.

Another recent study \cite{Bees00} focused on the swarming of {\em Serratia
liquefaciens}.  In that case the structure of the molecular feedback loops are
better explored, and were resolved in the model.  The production of a wetting
agent was initiated by high concentrations of specific 
signalling molecules. The colony expansion was considered to be a direct
consequence of the flow of the wetting fluid film, in which process the only
effect of bacteria (besides the aforementioned production) was changing the
effective viscosity of the fluid. The wetting agent production was
downregulated through a negative feedback loop involving swarmer cell
differentiation. This scenario is certainly not applicable to {\em P.
Mirabilis}, where swarmer cells actively migrate outwards and their role is
quite the opposite: enhancing the expansion of the colony.

The first theoretical analysis focusing on {\it P. mirabilis} was performed by
Esipov and Shapiro (ES) in \cite{ES98}. Their model was constructed based on
assumptions similar to ours, and could reproduce the alternating migration and
consolidation phases during the colony expansion. However, the complexity of
the ES model involves a rather large number of  model parameters, which
practically impedes both the full mapping of the parameter space and the
quantitative comparison of the model results with experimental findings.  
The major differences between our and the ES model can be summarized as
follows: (i) we do not resolve the {\em age} of the swarmer population.
Instead, we have a density measure and a constant decay rate implying an
exponential lifetime distribution on the (unresolved) level of individual
cells.  Since the available microbiological observations \cite{AH91,MTNIWM00}
suggest {\em only} that the lifetime is finite, there is no reason for
preferring any specific distribution.  (ii) We did not incorporate into our
model an unspecified ``{\em memory field}'' with a {\em built-in hysteresis}.
Instead, we implemented a {\em density-dependent motility} of the swarmer cells,
which behavior has been indeed observed \cite{WS78,AH91,FH99}.  (iii) In our
model the fully differentiated swarmer cells {\em do not grow}, which
assumption is probably not fundamental for the reported behavior, but it seems
to be more realistic because of the repression of many biosynthetic pathways
\cite{AH91}.  Finally, (iv) we do not consider any specific {\em interaction
between the motility of swarmer cells and the non-motile vegetative cell
population}.  Although such interactions probably exist, they are undocumented,
and as we demonstrated, are not required for the formation of periodic
swarming cycles.  However, such effects can be important in the actual
determination of the density profiles.

With these differences, which are not compromising the biological relevance of
the model, we were able to map {\em completely} the phase diagram,
establish approximate analytical formulas and estimate the value of {\em all}
model parameters in the case of four different strains.  In addition,
experimental data measured under various conditions could be explained with one
particular parameter setting in the case of the PRM1 strain indicating the
predictive power of our approach. Our model is a {\em minimal} model in the
sense that all of the explicitly considered effects (thresholds, diffusion,
etc.) were required to produce the oscillatory behavior, thus, it can not be
simplified further. Such minimal models can serve as a comparison baseline for
later investigations of various specific interactions.

The values of the microscopic parameters of the model can be either measured
directly (like $r_0$, $\rho^*_{min}$, $\rho_{min}$, $\rho_{max}$ or $r^*$) or
can be determined indirectly from experimental data (as $\rho_{min}^o$ and
$r$).  Most of these measurements have not yet been performed, we hope that our
work will motivate such experiments further examining the validity of
our assumptions. In fact, one of the parameters, $r^*/r_0$ was set to 1
during the fitting processes, as currently there is no available data to
estimate its value. Our numerical results suggest that it is probably larger
than $0.3$, and it is unlikely to be larger than 2 (meaning an average lifetime
less than 30 minutes). Within this range our qualitative conclusions are valid,
while the numerical values of the parameter estimates can change up to a factor
of $3$.

The behavior of ``precocious'' swarming mutants reported in \cite{BSM98}
deserves special attention. First, we would like to comment on the huge
difference found in the value of the transition rate $r$ (see Table II). We
emphasize that this is not an arbitrary output of a multiparameter fitting
process. First, we have reasons to believe, that the motility thresholds of the
two BB strains are rather similar. Knowing the growth rates and the cycle
times, Eq.  (\ref{resT}) shows us that the difference in the values of
$\rho_{min}$ can not exceed one order of magnitude. Assuming then this maximal
difference in $\rho_{min}$, $r$ remains the {\em only} free variable in Eqs.
(\ref{resL0})-(\ref{resL2}), and the fitting can be performed unambigously.
Thus, we are quite confident that such a large difference exists in $r$ showing
that the {\it rsbA} gene (in which these strains differ) influences not only
the cell density threshold, but the rate of differentiation as well. It is also
interesting to note that in Fig.~8b the behavior of the PRM2 and PRM1 strains
reflect a relation very similar to that of the BB2000 and BB2235 strains.
Finally, our calculations predicted a slightly longer cycle time for the
precocious swarming mutant BB2235, which is also in accord with the actual
experimental findings (see Fig~2. of \cite{BSM98}).

In our model
the assumed functional form of the density-dependence of the diffusion 
coefficient is somewhat different from the  most often considered one 
\cite{Golding98,Mimura00}, namely
\be
D(\rho)\sim\rho^k.
\label{nonlinD2}
\ee
The advantage of (\ref{nonlinD2}) is that it allows analytic solutions for 
certain
cases \cite{Murray}, however, it describes an unlimited, arbitrarily fast
diffusion inside the colony where the density is high. In contrast, in real
colonies the diffusion of cells is certainly bounded, and the expansion of the
boundary can be often limited by the supply of cells from behind
\cite{RMWSES96}.  Therefore we believe that our thresholded formulation
(\ref{nonlinD}) is a better approximation of what is taking place inside the
real colonies. 

Finally we would like to comment on the role of nutriens in the swarming
behavior of {\it P. mirabilis}. In our model there is a phenomenological 
parameter ($\tau$) determining how long the swarmer cells are produced at a
given position in the colony. When investigating the dependence of the cycle
time on this parameter, as Fig.~5  demonstrates, we found an extremely weak
effect.  Thus, at least within the framework of this model there is no
contradiction between the assumption that the swarmer cell {\em production} 
ceases due
to nutrient (or accumulated waste) limitations, and the seemingly
nutrient-independent cyclic behavior. In fact, this idea can be developed
further. By increasing $\tau$ (or decreasing the motility threshold
$\rho^*_{min}$) we arrive into a regime where the migration/consolidation phases
are not clearly separable as a motile swarmer cell population
exists even when the expansion of the colony is slower.
Experiments mapping the morphology diagram of {\it P. mirabilis} (Fig.~2 of
\cite{Nakahara96}) showed that there are certain values of agar hardness and
nutrient concentration, for which the expansion of the colony is still
oscillating, but the periodic density changes are smeared out due to the
presence of motile swarmer cells in the consolidation periods.  If one
associates the increasing agar hardness with increasing $\rho^*_{min}$ and the
nutrient concentration with $\tau$ then one can qualitatively reproduce those
(i.e., the $P_r$ and $P_h$) regions of the morphology diagram.

\acknowledgements 
One of the authors (M.M.) is grateful to T. Matsuyama and H. Itoh 
for many stimulating discussions on experimental results.  
This work was supported by funds OTKA T019299, F026645; FKFP 0203/197 and 
by grants No 09640471 and 11214205 from the Ministry of Education, Science 
and Culture of Japan.

\end{multicols}

\begin{table}
\begin{tabular}{|c|c||ccccc|c|c|c|}
Strain& 	& 		&     		& PRM1		&     		&     		& PRM2		& BB2000	& BB2235	\q
\hline
Experimental&
Temperature 	& 32$^o$C 	& 32$^o$C	& 32$^o$C	& 37$^o$C	& 22$^o$C	& 32$^o$C	& 37$^o$C	& 37$^o$C	\q
condition&
Agar 		& 2.0\%  	& 2.45\%	& 2.0\%		& n.a.		& n.a.		& 2.0\%		& n.a.		& n.a.		\q
&Reference 	&\cite{RMWSES96}&\cite{RMWSES96}& \cite{IWMM99} &\cite{RMWSES96}&\cite{RMWSES96}&\cite{RMWSES96}& \cite{BSM98}  & \cite{BSM98}	\q
\hline
&$T$ [h]  	& 4.7 		& 4.7      	& 4.0      	& 3.5      	& 8.5      	& 6.0      	& 3.0      	& 3.1      	\q
Colony-level&
$v$ [mm/h] 	& 1.7      	& 0.6       	& 1.0      	& n.a.		& n.a.		& n.a.		& 3.3      	& 3.3      	\q
&$w$ [mm] 	& 8.0      	& 3.0      	& 3.8      	& n.a.		& n.a.		& n.a.		& 10         	& 10 		\q
\hline
Cellular-level&
$r_0$ [1/h] 	& 0.6 		& 0.6		& n.a. 		& 1.0		& 0.4		& n.a. 		& n.a. 		& n.a. 		\q
\end{tabular}

\caption{
Summary of experimental data for four 
different {\it P. mirabilis} strains under various experimental conditions.
The value of $r_0$ was determined from growth monitoring in liquid cultures.
}
\end{table}

\begin{table}
\begin{tabular}{|c|c|ccccc|c|c|c|}
Strain&		& 		&     		& PRM1		&     		&     		& PRM2		& BB2000	& BB2235	\q
\hline
Experimental&
Temperature 	& 32$^o$C 	& 32$^o$C	& 32$^o$C	& 37$^o$C	& 22$^o$C	& 32$^o$C	& 37$^o$C	& 37$^o$C	\q
condition&
Agar 		& 2.0\%  	& 2.45\%	& 2.0\%		& n.a.		& n.a.		& 2.0\%		& n.a.		& n.a.		\q
\hline
Microscopic
&$r_0$ [1/h] 	& 0.53$^*$ (0.6)& 0.53$^*$ (0.6)& 0.7$^*$	& 1.0 		& 0.4		& 1.0$^*$  	& 2.5$^*$  	& 1.5$^*$  	\q
parameters
&$\rho_{min}$ [cell/$\mu$m$^2$] 
		&       	&       	& $0.06$	&       	&       	& $0.6$		& $2.0$		& $0.2$		\q
(independent)
&$\rho^*_{min}$ [cell/$\mu$m$^2$] 
		&       	&       	&$6\cdot10^{-3}$&       	&       	&$6\cdot10^{-3}$&$4\cdot10^{-3}$&$4\cdot10^{-3}$\q
&$\rho_{max}$ [cell/$\mu$m$^2$] 
		&       	&       	& $0.6$		&       	&       	& $240$		& $3000$	& $2.2$		\q
&$r/r_0$	&  		& 		& $10^{-1}$	& 		& 		& $10^{-4}$	&$2\cdot10^{-5}$&$3\cdot10^{-2}$\q
&$D_0$ [mm$^2$/h]& 20		& 3.2		& 6		& --		& --		& --		& 60		& 40		\q
\hline
(derived)
&$\rho^0_{min}$ [cell/$\mu$m$^2$] 
		&		&		&$3\cdot10^{-3}$& 		& 		&$3\cdot10^{-2}$& $10^{-1}$	& $10^{-2}$	\q
&$P/\rho_{min}$&  		& 		& 0.9		& 		& 		& 0.04		& 0.03 		& 0.3		\q
&$\tau/T$	&  		& 		& 0.7		& 		& 		& 0.9		& 0.9 		& 0.3		\q
&$\rho^*_{min}/\rho_{min}$
		& 	 	& 		& $10^{-1}$	& 		&  		& $10^{-2}$	&$2\cdot10^{-3}$&$2\cdot10^{-2}$\q
&$x_0$ [mm] 	& 6		& 2.3		& 3.0		& --		& --		& --		& 5.0		& 5.0		\q
&$v_0$ [mm/h] 	& 50 		& 20		& 27		& --		& --		& --		& 85		& 70		\q 
\hline
Macroscopic
&$T$ [h]  	& 5.5 (4.7) 	& 5.5 (4.7) 	& 4.7 (4.0)	& 3.3 (3.5) 	& 8.2 (8.5) 	& 5.6 (6.0) 	& 2.9 (3.0) 	& 3.2 (3.1) 	\q
behavior
&$v$ [mm/h] 	& 1.4 (1.7) 	& 0.5 (0.6)  	& 0.8 (1.0) 	& n.a.		& n.a.		& n.a.		& 3.4 (3.3) 	& 3.1 (3.3) 	\q
&$w$ [mm] 	& 7.8 (8.0) 	& 3.0 (3.0)	& 3.9 (3.8) 	& n.a.		& n.a.		& n.a.		& 10 (10)	& 10 (10)	\q
\end{tabular}

\caption{
Model parameters and the corresponding results for the strains and experimental
conditions specified in Table I. The model has seven microscopic parameters,
the rates $r_0$, $r$, $r^*$, the threshold densities $\rho_{min}$,
$\rho^*_{min}$ and $\rho_{max}$ and the diffusivity $D_0$. For each of the
strains $r^*=r_0$ was assumed. For comparison, other (derived) microscopic
parameters are also included. The calculated period lengths, terrace sizes and
expansion speeds are also presented together with the corresponding
experimental values (in parentheses).  The values marked by an asterisk ($*$)
were derived from the lag phase length data based on Eqs. (\ref{resL0}) and 
(\ref{resL2}).
Note the similarity between the PRM1 and BB2235, and also between the PRM2 and
BB2000 strains.
}
\end{table}

\begin{multicols}{2}

\end{multicols}

\end{document}